# Canonical Cortical Circuits and the Duality of Bayesian Inference and Optimal Control

Kenji Doya

doya@oist.jp

Neural Computation Unit, Okinawa Institute of Science and Technology Graduate University

1919-1 Tancha, Onna, Okinawa, 904-0495, Japan

**Highlights:**

- The duality of sensory inference and optimal control has recently been recognized as the commonality in computations for the posterior distributions in dynamic Bayesian inference and the value functions in reinforcement learning.

- We consider the hypothesis that the sensory and motor cortical circuits implement dual computations of sensory inference and optimal control, or perceptual and value-based decision making, respectively.

**Abstract:**

The duality of sensory inference and motor control has been known since the 1960s and has recently been recognized as the commonality in computations required for the posterior distributions in Bayesian inference and the value functions in optimal control. Meanwhile, an intriguing question about the brain is why the entire neocortex shares a canonical six-layer architecture while its posterior and anterior halves are engaged in sensory processing and motor control, respectively.

Here we consider the hypothesis that the sensory and motor cortical circuits implement the dual computations for Bayesian inference and optimal control, or perceptual and value-based decision making, respectively. We first review the classic duality of inference and control in linear quadratic systems and then review the correspondence between dynamic Bayesian inference and optimal control. Based on the architecture of the canonical cortical circuit, we explore how different cortical neurons may represent variables and implement computations.

## 1. Introduction

Sensory perception and motor control are the most fundamental functions of the brain. Although they are often studied separately, sensory perception and motor control are dependent on each other, which calls for an integrated approach. In addition to composing a sensory-motor loop, computations for sensory inference and optimal control have been shown to have a similarity. Rudolph Kalman [1] showed that the equations used for optimal sensory inference by the Kalman filter are similar to the equations used for optimal motor control by a linear quadratic regulator (LQR) [2]. This is known as Kalman's duality. More recently, researchers in reinforcement learning discovered more general correspondences between the computations for the posterior distribution in dynamic Bayesian inference and the value function in reinforcement learning [3-11]. This notion has created a research strategy called control as inference, or reinforcement learning as inference, which provides novel mathematical insights and helps the development of new reinforcement learning algorithms.

Regarding the brain's architecture for sensory perception and motor control, the most fundamental division is between the posterior half of the cerebral cortex, which is mostly involved in sensory perception, and the anterior half, which is mostly involved in motor control, planning and decisions. An interesting unanswered, or unasked, question is: Why does the entire cerebral cortex share the same canonical circuit architecture [12,13] characterized by a common six-layer architecture with specific types of neurons and connections for implementing sensory inference and motor control?

Here we consider the hypothesis that the sensory and motor cortical circuits evolved to make the dual computations for sensory inference and optimal control, or perceptual and value-based decision making, respectively. We first review the classic duality of inference and control in linear quadratic systems, known as Kalman's duality, and then review a more general correspondence between dynamic Bayesian inference and optimal control. We then explore how different types of cortical neurons in sensory and motor cortices may represent different variables and what cortical dynamics may realize required computations. We further discuss what experimental and computational approaches are required for scrutinizing this dual cortical circuit hypothesis.

## 2. The Duality of Control and Inference

Kalman filter [1] is the standard method for keeping track of the signal of interest despite noisy observation based on the assumption of linear dynamics and Gaussian noise (**Box 1**). Kalman pointed out that the set of equations for updating the estimates of the mean and the covariance of the state variable has the same structure as the equations for optimal control of a linear dynamical system with Gaussian noise [2]. This is known as Kalman's duality.

While Kalman's duality has been document in textbooks of control theory and signal processing as a matter of mathematical beauty, recent researchers in reinforcement learning found the relationship can extend beyond linear-Gaussian cases and have developed novel reinforcement learning and control algorithms based on the notion.

Emanuel Todorov pointed out the general duality between computations needed for posterior distributions in dynamic Bayesian inference and the value functions in optimal control and reinforcement learning (**Box 2**) [7-9]. A similar correspondence was also formulated by other researchers as well [3-6,10].

Based on this notion, Todorov realized that, by defining the cost of action as the divergence of the state transition probability from that by 'passive dynamics,' the exponentiated state value function can be computed linearly, which drastically reduces the required data and computation, and enables compositionality of value functions for different goal rewards [14,15]. A similar approach was also derived by Kappen [3,4].

Most recently, Sergey Levine reviewed all these works and formulated a probabilistic graphical model (PGM) having the optimality variable, which takes 1 if a state-action pair is optimal (**Figure 2**B) [11]. By assuming that the optimality variable follows a probability given by the exponential of a reward function, a standard message passing algorithm for PGM turns into the update equations for state and action value functions. For this conversion to hold, the objective function should include a regularization term for the entropy of action policy, which is known as maximum-entropy reinforcement learning [16].

**Table 2** summarizes the correspondence between the components of Bayesian inference and optimal control. The framework presents a unified theoretical basis for efficient and robust reinforcement learning algorithms, such as the soft actor-critic [17], and is expected to promote derivation of novel algorithms.

## 3. Canonical Cortical Circuits

The cerebral neocortex has a common six-layer architecture, known as the *canonical cortical circuit* (**Figure 3**) [12,13,18]. While most studies focused on the sensory cortex, the architecture of motor cortex with the thalamic inputs originating from the cerebellum and the basal ganglia has also been worked out [19]. A marked difference between the sensory and motor cortices is the thickness of the layer 4, which is densely populated by excitatory stellate cells in the sensory cortex. Despite a quantitative difference across areas, the basic architecture is preserved: layer 4 receives bottom-up thalamic and cortical input and projects to layers 2/3, where neurons have dense recurrent connections. Layer 2/3 pyramidal neurons project to higher cortical areas and also send output to layers 5/6. Layer 5 pyramidal neurons project to the cerebellum and the basal ganglia and layer 6 pyramidal neurons project to the thalamus and lower cortical areas.

Given the evidence suggesting that sensory perception is based on Bayesian inference combining top-down prediction and bottom-up sensory evidence [20-22], hypotheses have been proposed about how Bayesian inference can be implemented in the canonical cortical circuits [23-30]. Models of the motor cortex have also been proposed based on optimal control theory [31,32].

## 4. Dual Cortical Computation Hypothesis

Considered together, the duality of Bayesian inference and optimal control and the canonical cortical circuits in the sensory and motor areas suggest that common computations for inference and control are implemented in the common architecture of the neural circuits in the sensory and motor cortices, or the posterior and anterior halves of the cerebral cortex.

**Figure 3** and **Table 3** illustrate the canonical cortical circuits in the sensory and motor cortical areas and hypothetical representations of key variables of Bayesian inference and optimal control.

For sensory inference, the thalamic and cortical bottom-up inputs represent sensory observation $o_t$, which is used for evaluating the likelihood $p(o_t|s)$ of different world states $s$ in layer 4. The layer 2/3 combines the sensory likelihood with the predicted state $p(s_t|s_{t-1})$ to represent the surprise signal $p(o_t|s_t)$, which is sent to higher cortical areas. This information is also sent to layer 5/6 to update the posterior probability $p(s_t|o_1, ..., o_t)$. These computations may also be conditional on the top-down contextual signal $z_t$, including the executed action $a_{t-1}$, from higher cortical areas.

In optimal control, the reward function $r(s, a)$ and the state value function $V(s)$ correspond to the log likelihood and log posterior probability in sensory inference, so that they would be represented by layer 4 and layer 5/6 neurons, respectively. The update of action value function $Q(s, a)$ requires state transition model $p(s_{t+1}|s_t, a_t)$ and reward information, which is likely to be represented by layer 2/3 neurons. The action policy $p(a|s)$ is computed by subtracting the state value from the action value, so that action may be selected in layer 5 or 6 and sent to lower cortical and subcortical areas. Note that the above is just one hypothetical realization and many other mappings of different roles to neurons and connections are conceivable.

There are many interesting open questions about the cortical implementation of the dual computations for Bayesian inference and optimal control. First, how the backward computation is realized in real time? In the visual cortex, evidence suggests that the alpha rhythm around 10 Hz carries top-down feedback information [33] and underlies multi-modal sensory arbitration [34]. In the motor cortex, the beta rhythm around 20 Hz shows responses before execution or during imagination of movements [35,36]. These might be the correlates of periodic execution of backward computation.

Another important question is how the state transition model $p(s_{t+1}|s_t, a_t)$ and the sensory observation model $p(o_t|s_t)$ are learned, together with the internal representations of state $s$ and action $a$. The roles of the cerebellar and the basal ganglia inputs through the thalamus to the motor cortex in learning is also an interesting question [37,38].

Finally, how are the parameters for Bayesian inference and optimal control regulated, such as the time frame of planning and the prior uncertainty of the state dynamics and sensory observation? The roles of neuromodulators, such as serotonin, noradrenaline and acetylcholine have been suggested [39-44].

Given recent advances in two-photon calcium imaging [45,46] and electrode array recording [47], it is now feasible to test such hypotheses regarding the implementation of Bayesian inference and optimal control by large-scale measurement of neural activities during sensorimotor tasks [38,48,49]. The correspondence between variables and mappings for Bayesian inference and optimal control as depicted in **Table 3** may provide a basis for interpreting data from both sensory and motor cortices and coming up with a unified theory of cortical computation.

## 5. Conclusion

This article reviewed the duality between sensory inference and motor control and presented a hypothesis that the canonical posterior and anterior cortical circuits perform such dual computation. The author believes that this overreaching hypothesis is worthy of experimental testing by utilizing multi-area, multi-layer neural recording technologies. In addition to the basic operations for inference and control, the regulatory mechanisms for such computations and possible malfunctioning of such mechanisms would provide better understanding of the roles of neuromodulators and the causes of psychiatric disorders.

**Box 1: Kalman's duality.** The computations for optimal filtering and optimal control under linear Gaussian assumptions reduces to solving the same form of matrix equations [1,7].

**Figure 1: A. Kalman filter.** We consider a linear discrete-time dynamical system

$$x_{t+1} = Ax_t + Bu_t + v_t \quad (1)$$

$$y_t = Cx_t + w_t \quad (2)$$

where $x_t$ is the state, $u_t$ is the action input, $y_t$ is the sensory observation, and $v_t$ and $w_t$ are the state and observation noises with covariance matrices $S$ and $U$, respectively. We aim to estimate the changing state $x_1, \ldots, x_t$ iteratively from the observations $y_1, \ldots, y_t$. We represent the uncertainty of the state by a Gaussian distribution $x_t \sim \mathcal{N}(\bar{x}_t, \Sigma_t)$.

With the state transition (1), the mean and the variance of the state distribution evolve as

$$\hat{x}_{t+1} = A\bar{x}_t + Bu_t \quad (3)$$

$$\hat{\Sigma}_{t+1} = S + A\Sigma_t A' \quad (4)$$

where $A'$ is the transpose matrix of $A$. With a new observation $y_{t+1}$, the distribution is updated by Bayesian inference with the predicted distribution $\mathcal{N}(\hat{x}_{t+1}, \hat{\Sigma}_{t+1})$ as the prior and the observation giving the likelihood $\mathcal{N}(y_{t+1} - C\hat{x}_{t+1}; 0, U)$. This leads to an update of the mean of the state distribution in proportion to the prediction error

$$\bar{x}_{t+1} = \hat{x}_{t+1} + K_t(y_{t+1} - C\hat{x}_{t+1}) \quad (5)$$

where the update gain is given by

$$K_t = A\Sigma_t C'(U + C\Sigma_t C')^{-1} \quad (6)$$

This is called the filter gain, or Kalman gain, which becomes large when the state uncertainty $\Sigma(t)$ is large. The variance of the state distribution is also updated by the Kalman gain as

$$\Sigma_{t+1} = \hat{\Sigma}_{t+1} - K_t C\Sigma_t A' \quad (7)$$

which generally reduces the uncertainty. Taken together (4), (6) and (7), the update of the state covariance is given as

$$\Sigma_{t+1} = S + A\Sigma_t A' - A\Sigma_t C'(U + C\Sigma_t C')^{-1} C\Sigma_t A' \quad (8)$$

**B: Linear quadratic regulator (LQR).** Here we consider the same dynamical system (1) and consider the cost function (negative reward)

$$l(x, u) = \frac{1}{2} x'Qx + \frac{1}{2} u'Ru \quad (9)$$

where $Q$ and $P$ are matrices defining the state and action costs [2]. The aim is to minimize the cumulative cost

$$J = \sum_{t=1}^{T} l(\boldsymbol{x}_t, \boldsymbol{u}_t) \tag{10}$$

In the linear Gaussian assumption, the state value function $V(\boldsymbol{x}, t)$ can be represented by a quadratic form of the state vector $\boldsymbol{x}$ and a matrix $P_t$ as

$$V(\boldsymbol{x}, t) = \frac{1}{2}\boldsymbol{x}'P_t\boldsymbol{x} = E\left[\sum_{s=t}^{T} l(\boldsymbol{x}_s, \boldsymbol{u}_s) \,|\, \boldsymbol{x}_t = \boldsymbol{x}\right] \tag{11}$$

The optimal action is given by the Bellman equation

$$\frac{1}{2}\boldsymbol{x}'P_t\boldsymbol{x} = \min_{\boldsymbol{u}}\left[l(\boldsymbol{x}, \boldsymbol{u}) + \frac{1}{2}(A\boldsymbol{x} + B\boldsymbol{u})'P_{t+1}(A\boldsymbol{x} + B\boldsymbol{u})\right] \tag{12}$$

The solution leads to a feedback control law

$$\boldsymbol{u}(t) = -G_t \boldsymbol{x}_t \tag{13}$$

where the feedback control gain is given by

$$G_t = (R + B'V_{t+1}B)^{-1}B'P_{t+1}A \tag{14}$$

The matrix for the value function is computed backward in time

$$P_t = Q + A'P_{t+1}A - A'P_{t+1}BG_t \tag{15}$$

with the terminal condition of $P_T = Q$. From (14) and (15), the update equation for $V$ is

$$P_t = Q + A'P_{t+1}A - A'P_{t+1}B(R + B'P_{t+1}B)^{-1}B'P_{t+1}A \tag{16}$$

which has a similar form with the equation (8) for Kalman filter.

The table below summarizes the correspondence between Kalman filter and LQR [1,7,8].

**Table 1:** The duality of the Kalman filter [1] and the linear quadratic regulator (LQR) [2].

| Kalman Filter | Linear Quadratic Regulator |
|---|---|
| covariance of state distribution $\Sigma_t$ | coefficients of quadratic state value $P_t$ |
| filter gain | control gain |
| $K_t = A\Sigma_t C'(U + C\Sigma_t C')^{-1}$ | $G_t = (R + B'P_{t+1}B)^{-1}B'P_{t+1}A$ |
| state dynamics $A$ | transpose state dynamics $A'$ |
| observation $C$ | action sensitivity $B'$ |
| state noise covariance $S$ | state cost coefficients $Q$ |
| observation noise covariance $U$ | action cost coefficients $R$ |
| forward from $\Sigma_1 = S$ | backward from $P_T = Q$ |

**Box 2: Duality of Bayesian inference and optimal control.**

Instead of just formal similarity, there is a meaningful correspondence between dynamic Bayesian inference and optimal control or reinforcement learning [7-9].

**Figure 2: A. Dynamic Bayesian inference.** Here we consider the stochastic state dynamics

$$p(s_{t+1}|s_t, a_t) \tag{1}$$

and the sensory observation

$$p(o_t|s_t) \tag{2}$$

In active perception, the aim is to estimate the state trajectory behind a series of actions and observations $p(s_1, \ldots, s_T | a_1, o_1, \ldots, a_T, o_T)$. In observation learning, the sequence of actions, such as muscle forces, are estimated from observation as $p(s_1, a_1, \ldots, s_T, a_T | o_1, \ldots, o_T)$.

Here we outline the simplest case involving no action, $p(s_{t+1}|s_t, a_t) = p(s_{t+1}|s_t)$. This is an example of a hidden Markov model (HMM) and a standard algorithm for solving the problem is the forward-backward algorithm [50,51]. We define the forward message as the joint probability of the observations up to $t$ and the state $s_t$

$$\alpha(s_t) = p(o_1, \ldots, o_t, s_t) \tag{3}$$

and the backward message as the conditional probability of observations after the state $s_t$

$$\beta(s_t) = p(o_{t+1}, \ldots, o_T | s_t) \tag{4}$$

The forward and backward messages are iteratively computed as

$$\alpha(s_t) = p(o_t|s_t) \int_S \alpha(s_{t-1}) p(s_t|s_{t-1}) ds_{t-1} \tag{5}$$

$$\beta(s_t) = \int_S \beta(s_{t+1}) p(o_{t+1}|s_{t+1}) p(s_{t+1}|s_t) ds_{t+1} \tag{6}$$

The posterior probability is then computed from the product of the two messages.

$$p(s_t|o_1, \ldots, o_T) \propto \alpha(s_t)\beta(s_t) \tag{7}$$

**B. Optimal control as inference.** There are several ways to cast optimal control problems into dynamic Bayesian inference problems, which facilitate common understanding and development of new algorithms [3-10].

One way is to consider an "optimality variable $\mathcal{O}_t$" [11] which takes 1 when the state-action pair is optimal and assume that the reward function represents the log probability for the state-action pair to be optimal

$$p(\mathcal{O}_t = 1|s_t, a_t) = \exp(r(s_t, a_t)) \tag{8}$$

In this formulation, the posterior probability of state-action trajectory conditioned on the optimality variable to be 1 is given as

$$p(\mathbf{s}_1, \mathbf{a}_1, \ldots, \mathbf{s}_T, \mathbf{a}_T | \mathcal{O}_1 = 1, \ldots, \mathcal{O}_T = 1) \propto p(\mathbf{s}_1) \prod_{t=1}^{T} p(\mathcal{O}_t = 1 | \mathbf{s}_t, \mathbf{a}_t) p(\mathbf{s}_{t+1} | \mathbf{s}_t, \mathbf{a}_t)$$

$$= \left\{ p(\mathbf{s}_1) \prod_{t=1}^{T} p(\mathbf{s}_{t+1} | \mathbf{s}_t, \mathbf{a}_t) \right\} \left\{ \exp \sum_{t=1}^{T} r(\mathbf{s}_t, \mathbf{a}_t) \right\} \quad (9)$$

which means that state-action trajectories that is feasible under the state dynamics with higher cumulative rewards have high posterior probability.

This posterior probability can be computed by backward message passing. We define backward messages representing the state-action pair or the state is optimal for time *t* and future

$$\beta_t(\mathbf{s}_t, \mathbf{a}_t) = p(\mathcal{O}_1 = 1, \ldots, \mathcal{O}_T = 1 | \mathbf{s}_t, \mathbf{a}_t) \quad (10)$$

$$\beta(\mathbf{s}_t) = p(\mathcal{O}_1 = 1, \ldots, \mathcal{O}_T = 1 | \mathbf{s}_t) = \int_{\mathcal{A}} \beta(\mathbf{s}_t, \mathbf{a}_t) p(\mathbf{a}_t | \mathbf{s}_t) d\mathbf{a}_t \quad (11)$$

where $p(\mathbf{a}_t | \mathbf{s}_t)$ is a policy prior, such as uniform action selection. These messages are computed backward in time as

$$\beta_t(\mathbf{s}_t, \mathbf{a}_t) = \int_{\mathcal{S}} \beta_{t+1}(\mathbf{s}_t) p(\mathbf{s}_{t+1} | \mathbf{s}_t, \mathbf{a}_t) \exp(r(\mathbf{s}_t, \mathbf{a}_t)) d\mathbf{s}_{t+1} \quad (12)$$

Then the optimal policy is derived from these messages as

$$p(\mathbf{a}_t | \mathbf{s}_t, \mathcal{O}_t = 1, \ldots, \mathcal{O}_T = 1) \propto \frac{\beta_t(\mathbf{s}_t, \mathbf{a}_t)}{\beta_t(\mathbf{s}_t)} \quad (13)$$

These messages have correspondence with the value functions as

$$Q(\mathbf{s}_t, \mathbf{a}_t) = \log \beta_t(\mathbf{s}_t, \mathbf{a}_t) \quad (14)$$

$$V(\mathbf{s}_t) = \log \beta_t(\mathbf{s}_t) \quad (15)$$

under the addition of entropy-based regularization of the policy.

Table 2: The correspondence of dynamic Bayesian inference and optimal control.

| Bayesian inference | Optimal control |
|---|---|
| posterior state distribution $p(\mathbf{s}_t | \mathbf{o}_1, \ldots, \mathbf{o}_T)$ | state value function $\exp(V(\mathbf{s}_t))$ |
| state dynamics $p(\mathbf{s}_{t+1} | \mathbf{s}_t, \mathbf{a}_t)$ | state dynamics $p(\mathbf{s}_{t+1} | \mathbf{s}_t, \mathbf{a}_t)$ |
| observation model $p(\mathbf{o}_t | \mathbf{s}_t)$ | reward function $p(\mathcal{O}_t = 1 | \mathbf{s}_t, \mathbf{a}_t) = \exp(r(\mathbf{s}_t, \mathbf{a}_t))$ |
| forward message $\alpha(\mathbf{s}_t)$ | – |
| backward message $\beta(\mathbf{s}_t)$ | action value function $\exp(Q(\mathbf{s}_t, \mathbf{a}_t))$ |

**Figure 3:** Canonical cortical circuits in sensory and motor cortices and a hypothetical realization of dynamic Bayesian inference and optimal control. **A.** Possible realization of dynamic Bayesian inference in the sensory cortex. **B.** Possible realization of optimal control in the motor cortex.

**Table 3**: Correspondences of dynamic Bayesian inference and optimal control, and their possible implementation in the canonical cortical circuit.

| Inference | Cortex | Control |
|---|---|---|
| top-down signal $z_t$ | L1 input | top-down activation signal |
| bottom-up signal $p(o_t|s_t)$ | L2/3 output | action value $Q(s, a)$ |
| predictive model $p(s_t|s_{t-1})$ | L2/3 connection | predictive model $p(s_{t+1}|s_t, a_t)$ |
| bottom-up signal $o_t$ | L4 input | optimality signal $\mathcal{O}_t$ |
| likelihood $p(o_t|s)$ | L4 output | reward function $r(s, a)$ |
| posterior $p(s_t|o_1, \ldots, o_t)$ | L5 output | state value $V(s)$ |
| top-down signal $s_t$ | L6 output | action $p(a_t|s_t)$ |

**Acknowledgements**

This work was supported by JSPS KAKENHI Grants 23120007, 16K21738, 16H06561 and 16H06563, AMED Grants JP19dm0207001h0006 and JP21dm0307009, and research support of Okinawa Institute of Science and Technology Graduate University to KD. The author thanks Takuya Isomura and the reviewer for their comments on the manuscript.

**Annotations**

**[11] A tutorial article on the correspondence of probabilistic inference and optimal control and reinforcement learning.

** [17] One of the most successful reinforcement learning algorithms derived with the concept of control as inference.

*[46] A recent two-photon imaging system allowing recording from 3mm square cortical areas.

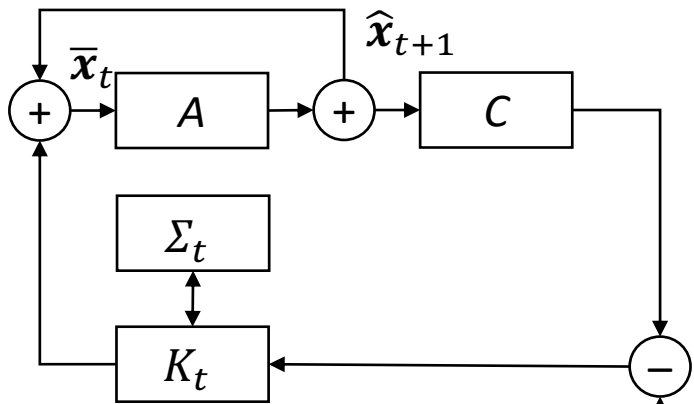
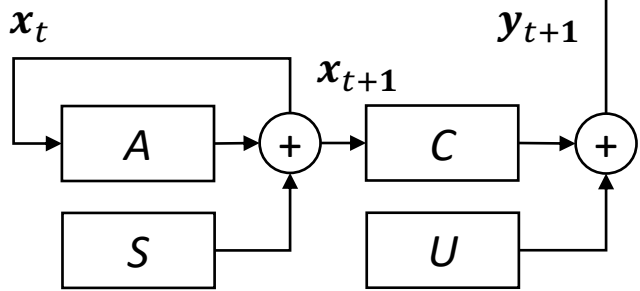
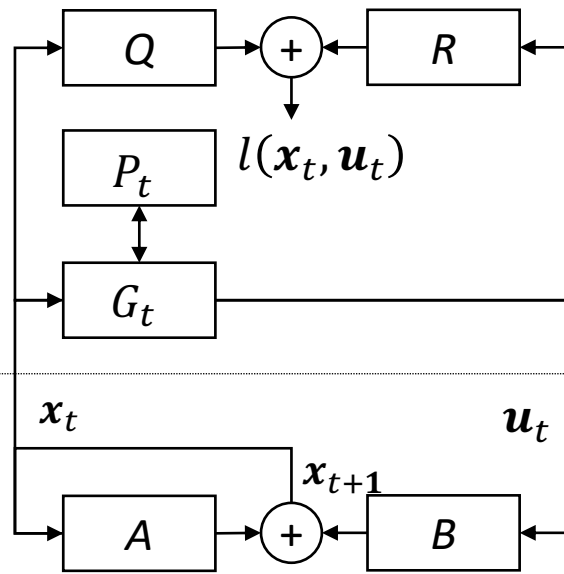

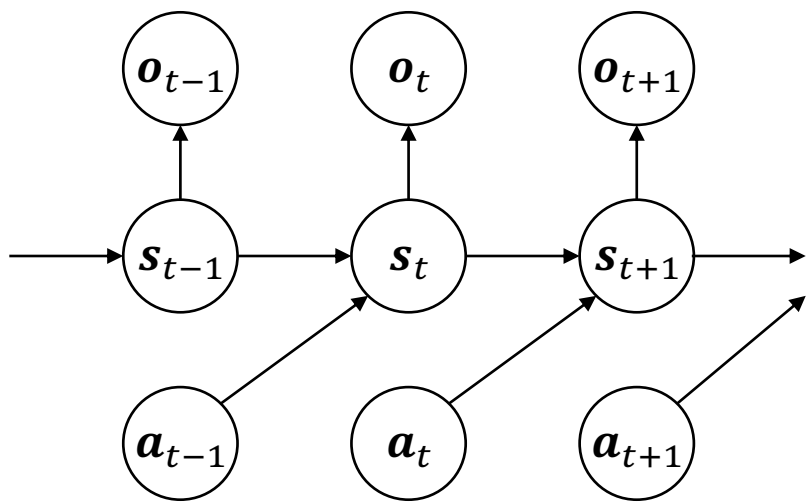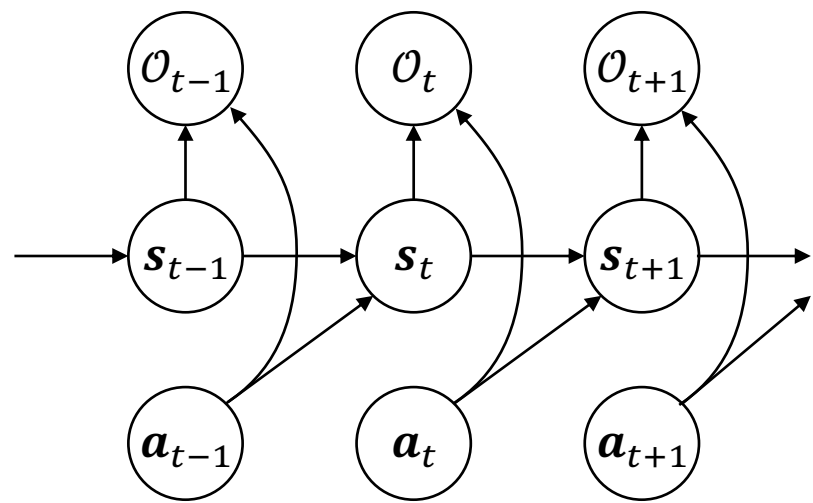

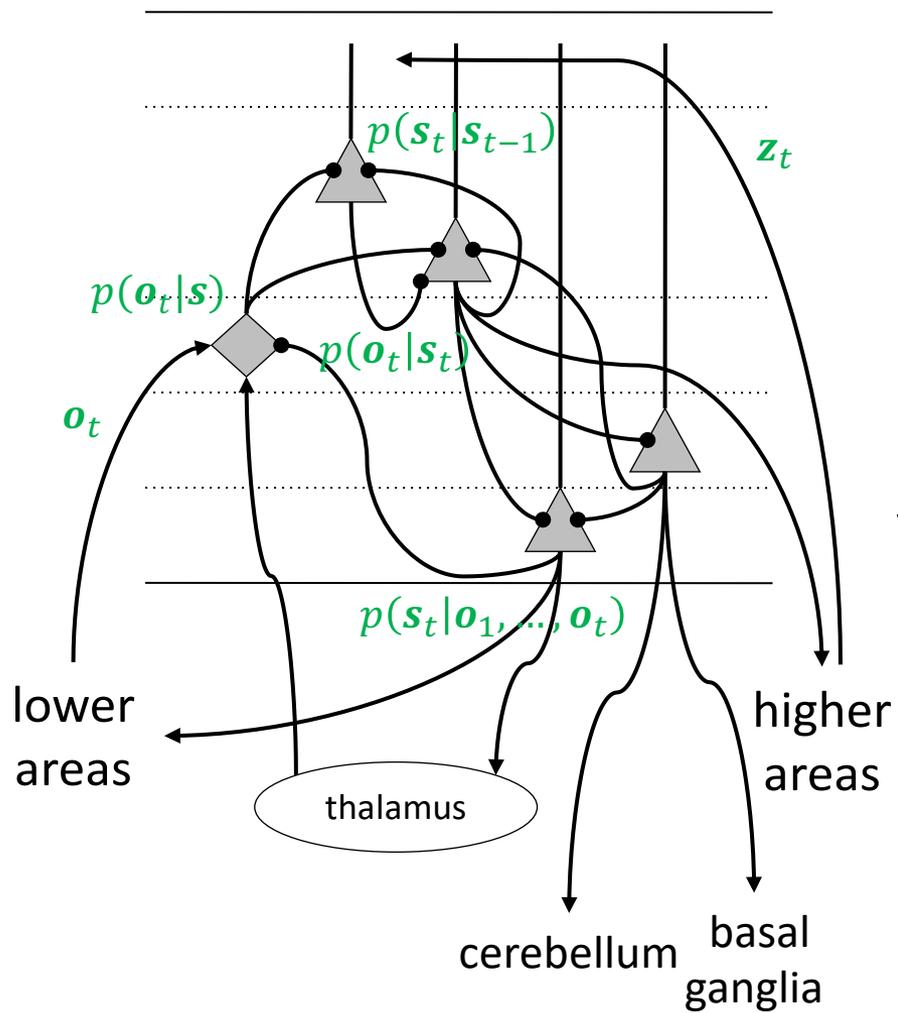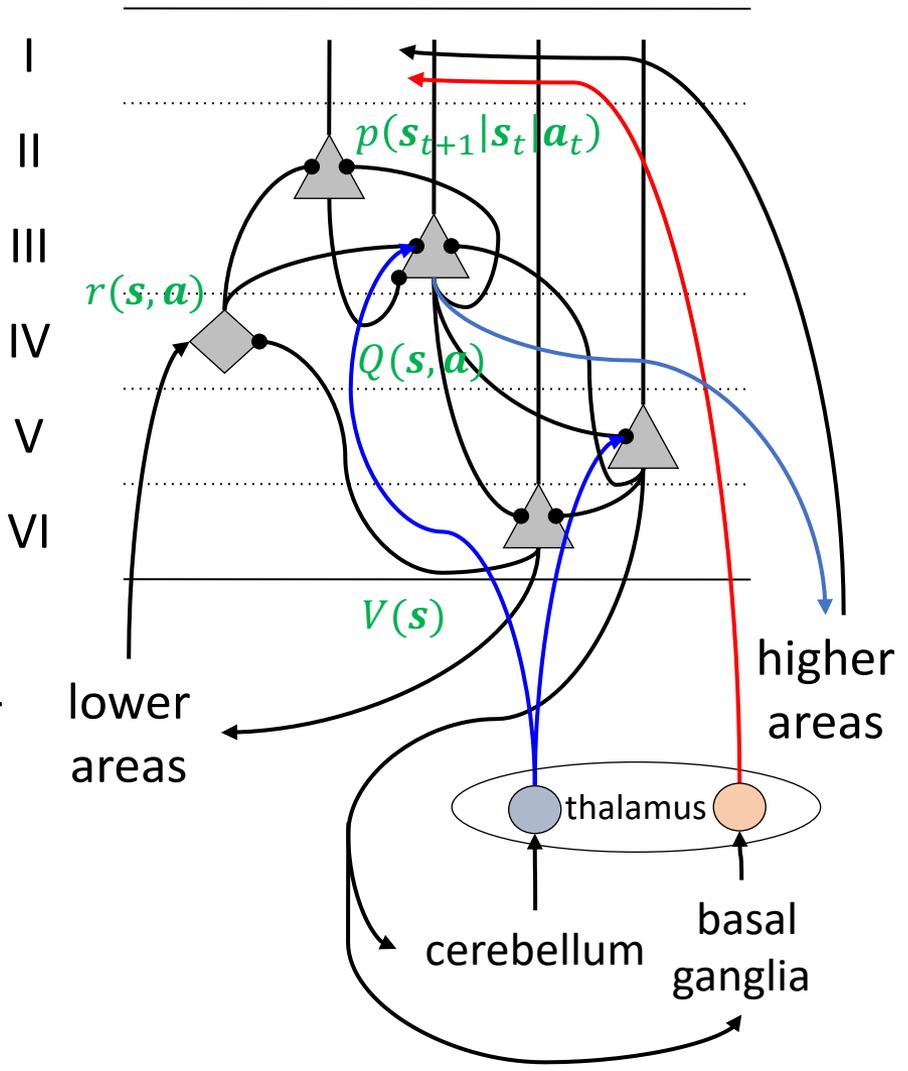